\documentclass[pdflatex,sn-nature]{sn-jnl}%

\usepackage{graphicx}%
\usepackage{multirow}%
\usepackage{amsmath,amssymb,amsfonts}%
\usepackage{amsthm}%
\usepackage{mathrsfs}%
\usepackage[title]{appendix}%
\usepackage{xcolor}%
\usepackage{textcomp}%
\usepackage{manyfoot}%
\usepackage{booktabs}%
\usepackage{algorithm}
\usepackage{algorithmicx}%
\usepackage{algpseudocode}%
\usepackage{listings}%
\usepackage[left]{lineno}
\usepackage{mdframed} 
\usepackage{natbib}

\definecolor{light-gray}{gray}{0.95}
\definecolor{commentgreen}{RGB}{2,112,10}
\definecolor{classiqpurple}{HTML}{bb8bff}
\definecolor{classiqred}{HTML}{F43764}
\definecolor{weborange}{RGB}{255,165,0}
\definecolor{frenchplum}{RGB}{129,20,83}
\begin{document}

\title[Design and synthesis of scalable quantum programs]{Design and synthesis of scalable quantum programs}

\author*{Tomer Goldfriend,
Israel Reichental,
Amir Naveh,
Lior Gazit,
Nadav Yoran,
Ravid Alon,
Shmuel Ur,
Shahak Lahav,
Eyal Cornfeld,
Avi Elazari,
Peleg Emanuel,
Dor Harpaz,
Tal Michaeli,
Nati Erez,
Lior Preminger,
Roman Shapira,
Erik Michael Garcell,
Or Samimi,
Sara Kisch,
Gil Hallel,
Gilad Kishony,
Vincent van Wingerden,
Nathaniel A. Rosenbloom,
Ori Opher,
Matan Vax,
Ariel Smoler,
Tamuz Danzig,
Eden Schirman,
Guy Sella,
Ron Cohen,
Roi Garfunkel,
Tali Cohn,
Hanan Rosemarin,
Ron Hass,
Klem Jankiewicz,
Karam Gharra,
Ori Roth,
Barak Azar,
Shahaf Asban,
Natalia Linkov,
Dror Segman,
Ohad Sahar,
Niv Davidson,
Nir Minerbi,
and Yehuda Naveh \sur{}}\email{tomer@classiq.io}

\affil{\orgname{Classiq Technologies}}
\affil{\orgaddress{\street{3 Daniel Frisch Street}, \city{Tel Aviv-Yafo}, \postcode{6473104}, \country{Israel}}}

\abstract{ 
We present a scalable, robust approach to creating quantum programs of arbitrary size and complexity. The approach is based on the true abstraction of the problem. The quantum program is expressed in terms of a high-level model together with constraints and objectives on the final program. Advanced synthesis algorithms transform the model into a low-level quantum program that meets the user's specification and is directed at a stipulated hardware. This separation of description from implementation is essential for scale. The technology adapts electronic design automation methods to quantum computing, finding feasible implementations in a virtually unlimited functional space. The results show clear superiority over the compilation and transpilation methods used today. We expect that this technological approach will take over and prevail as quantum software become more demanding, complex, and essential.
}

\maketitle

{\bf Quantum algorithms are algorithms designed to work on digital quantum computers. Most modalities of quantum computers comply with a set of logical, universal gates, from which any algorithm can be assembled~\cite{nielsen2010quantum}. 
Implementing a specific quantum algorithm on a specific quantum hardware, requires maturity of the quantum software stack, including design languages, compilers, and technology to map the logical description into a concrete, efficient implementation. As quantum computers grow in power and resources, the problem of efficiently mapping logical algorithms to concrete implementations becomes increasingly difficult and must advance from the manual methods existing today~\cite{Heim2020a}.
In this paper, we present a new approach, radically different from regular compilation methods~\cite{Qsharp2018,Qiskit2024,Pennylane}, to designing and implementing quantum algorithms. The approach, based on well-proven electronic design automation (EDA) principles and methods~\cite{wang2009electronic}, is shown to reduce required resources (e.g. qubit count, 2-qubit gate count) by orders of magnitude compared to other existing approaches. This is done by automatically managing resources such as available qubits, multiple function implementations, functional levels of accuracy, and other high-level logic equivalences.  Moreover, and in contrast to all existing approaches today, this approach is shown to be scalable to large-scale, realistic algorithms.
Following the path of classical design automation, we anticipate that a significant design gap will follow the increasing power of quantum computers. Building powerful and scalable software technologies for quantum algorithm design is an essential step towards any meaningful quantum advantage. This work presents the most advanced steps taken in this direction to date, and lays the foundation for the future of the quantum software stack.}

\section{Introduction}\label{sec:intro}

Since 2017, we have seen massive and continuous advances in quantum hardware in many modalities, including superconducting devices~\cite{Kandala_2017,GoogleSycamore,Krantz_2019,PRXQuantum.4.010314}, ion traps~\cite{Bruzewicz2019,IonQ2019,Pino_2021}, neutral atoms~\cite{Henriet2020quantumcomputing,Bluvstein_2023}, silicon devices~\cite{He2019ATG,Gonzalez_Zalba_2021}, photonic devices~\cite{Bartolucci2021FusionbasedQC,Madsen2022QuantumCA}, and more~\cite{doi:10.1126/science.abb2823}. The consensus is that quantum computing has shifted from a scientific/academic phase into an engineering phase during this time frame. The different hardware roadmaps vary, but all show major applicative breakthroughs that are expected in the next few years~\cite{IBM_roadmap,Google_roadmap,Quantinuum_roadmap,Ionq_roadmap,Quera_roadmap,Iqm_roadmap,Infleqtion_roadmap,Alice_and_bob_roadmap}.

Software concepts and platforms have not evolved at the same pace. The first software platforms~\cite{Qiskit2024,pytket_2020,Cirq,Quil} appeared hand-in-hand with the public exposition of quantum hardware accessible on the cloud in 2016~\cite{IBM_chip_on_cloud}. These platforms typically consisted of a means to specify the operations of quantum gates (and later quantum pulses) on a set of qubits, transpilation of the gate-level programs into equivalent gate-level programs suitable to a given hardware, and software for simulating the quantum hardware on classical CPUs and GPUs. Advancements in those software offerings included high-level constructs such as quantum registers and algorithmic building blocks. However, those were not true abstractions of the problem because the constructs only encapsulated hard-coded, low-level implementations of qubits and quantum gates; these were packaged as named-functions performing operations on named registers.

Further developments in quantum software included application libraries built on top of the basic software, including applications in quantum machine learning, finance, chemistry, and many more~\cite{Qiskit2024,Pennylane,Forest,Qiskit-nature}. These libraries provided further domain-specific algorithms and functions, based on building blocks tuned and optimized for a specific application. Fundamentally, these were not different from hard-coded quantum circuits encapsulated in several levels of building block functions. In addition, several quantum programming languages that describe functionality at a higher level of abstraction in different respects have been proposed in recent years~\cite{Qsharp2018,CudaQ,Pennylane_Catalyst}. The tool chains for these languages utilize compilation frameworks and techniques adapted from classical programming language processing, notably MLIR~\cite{MLIR} and LLVM~\cite{LLVM}. The actual compilation approach in all these cases still rigidly maps the abstractions onto a low-level implementation, taking only local resource considerations into account.

We present a radically different approach to the compilation of quantum programs. Our strategy follows and extends the general ideas presented earlier~\cite{wille18computer,wille20efficient,Heim2020a}. However, for the first time the ideas are fully implemented and do not remain at the conceptual levels. The level of understanding and detail gained through implementation clearly changes and cements the conceptual level as well; hence the work presented here creates a major step forward in advancing the path to true, scalable quantum software.

At the base of our work is abstraction. By abstraction we mean allowing the user to express any information of interest to them regarding the quantum program, and omit any other information, or detail, that is not of interest. It is up to the technology to fill in all details that were not expressly specified, and come up with an executable, optimal, quantum program that conforms to the specification of the user and fits a given hardware. The abstract description from the user is a \textit{model} of the program, while the outcome executable code is a \textit{concrete instance} of the model. In this paper, we will focus on functional models for which the abstract description describes required functionality of the program, constraints on the hardware, and possible optimization criteria.

At this level of description, there are many details that need to be filled. For example, each function may have many types of implementation, and each type can be implemented in itself in many different ways. An adder for example may be implemented as a simple ripple-carry adder, or as a Fourier transformed adder in which computation is done in quantum Fourier space. The result of the adder can be put out of place on a newly allocated register of qubits, leaving the two input registers intact, or can replace one or both of the input variables. Each operation within the adder may use any number of auxiliary qubits ('\textit{auxiliaries}') in order to perform the operation faster or with fewer two-qubit gates. Finally, the adder may be implemented to any level of accuracy, again affecting the number of qubits and two-qubit gates that the operation will consume. In all those cases, the functionality prescribed by the user remains intact\footnote{Playing with the accuracy degree of freedom results in different functionalities of the adder, however we consider a likely model in which only the accuracy of the entire program is specified, and not the accuracy of each component.}.

Beyond the single function, unspecified details may also relate to the interaction between functions. For example, different functions may or may not share auxiliary or functional qubits, depending on the scope of the function and whether its results are required downstream, and also on the cost of disentangling and resetting used qubits. More generally, the decision on which qubits, if at all, to disentangle and reuse, as well as the ordering of functions using the same auxiliary qubits, is a very important 'detail' from the implementation point of view, but is of no concern to the user from a functional model point of view.

Such considerations of 'filling in the details' between wide levels of abstraction are the basis of many electronic design automation (EDA) technologies, specifically synthesis from high-level functional and nonfunctional models to register-transfer-level (RTL) models and then to logic, netlist, and physical-level implementations. Given the circuit model of quantum computation, it should not be surprising that this general synthesis approach may also be applicable in the quantum domain. In this paper, we present the application of the EDA synthesis approach to quantum computing. Throughout the paper, we use the word \textit{synthesis} as its meaning in the EDA industry, which is much more general than the term sometimes used in quantum computing~\cite{F_rrutter_2024} - meaning the decomposition of a specific unitary into a given gate set. Our results below demonstrate the viability of this approach for creating robust quantum software at scale.

When developing the technology, we tried to preserve as much as possible all the guidelines and best practices of the conventional principles of EDA synthesis. The challenges we encountered in applying those principles to the nature of quantum programs were already significant enough so that we did not attempt to invent any new wheel when it was not necessary. Still, the peculiarities of quantum programs, including no fan-in or fan-out, no copy, disentanglement as means for freeing qubit resources, long-range correlations, inability to locally simulate sub-circuits, and many more -- make the incorporation of these concepts into EDA synthesis principles non-trivial and challenging. 

In this paper, we mainly focus on the synthesis technology aspect per se. While we present modeling examples when needed, we do not delve into the research behind the definition of the quantum modeling constructs and the common quantum idioms that we found leading to those constructs. In addition, we do not describe here the challenges behind analysis and debugging of the automatically synthesized quantum program, and feedback into improving the model from this analysis. The modeling constructs and automatic program analysis are important parts of the EDA flow that we have again adapted into the quantum software cycle but are beyond the scope of this paper.

\section{Results}\label{sec:results}

We present results showing the power of our approach, in which a truly abstract description implies functional flexibility that brings about optimized gate-level quantum programs. We use Qmod~\cite{Qmod_docs} and the Classiq open library~\cite{classiq_library} to model the abstract functionality of the program and synthesize concrete programs under hardware constraints and optimization functions using the synthesis engine, as described in the Methods section \ref{sec:methods} and the Supplementary Materials. We compare our results to those achieved by today's state-of-the-art quantum coding tools. The problems we have chosen for demonstration are the most generic one can think of, and the improvement in the results compared to state-of-the-art tools is only due to the new principles introduced by our approach.

\subsection{Quantum walk on a circle}
\label{sec:quantum_walk}

\begin{figure}[h!]
\begin{minipage}{\hsize}
\centering
\includegraphics[width=0.8\linewidth]{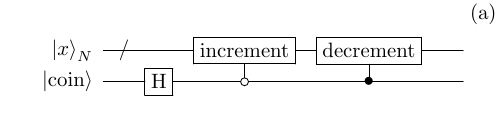}
\end{minipage}
\includegraphics[width=0.48\linewidth]{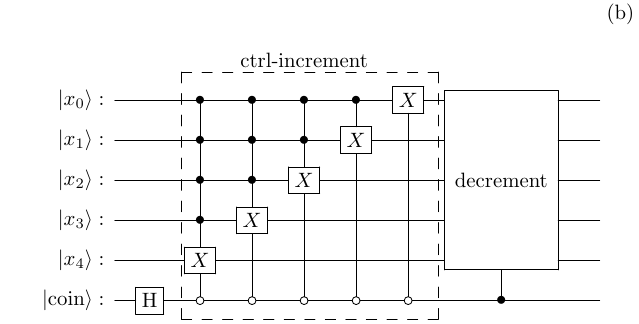}
\includegraphics[width=0.48\linewidth]{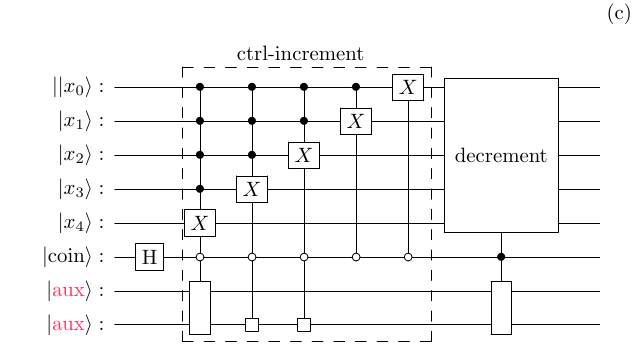}
\caption{Quantum walk on a circular graph with $2^N$ nodes. (a) A quantum circuit diagram for a single walking step, defined by three operations: a Hadamard gate on the coin qubit, followed by increment and decrement on the position qubit array, controlled on the coin qubit being at state 0 and 1, respectively. (b, c): Two functionally-equivalent, different implementations of the model. In both examples the controlled-increment block is opened one hierarchy down, showing its explicit implementation. (b): The result of synthesizing the functional model with width optimization, giving a quantum circuit of 5 qubits and CX-counts of 274.  (c): The result of synthesizing the functional model with CX-count optimization and constraining the width to 10 qubits. The synthesis outcome in this case is a quantum circuit of 7 qubits and CX-count of 120.}
\label{fig:increment_implementations}
\end{figure}

Initially inspired by random classical walks, quantum walk~\cite{10.1145/380752.380758} is an algorithmic technique that underlies many quantum algorithms, such as quantum search~\cite{QuantumSearch2007, Ambainis2007} and Hamiltonian simulation based on Qubitization~\cite{Berry_etal2015}. We consider the pedagogical example of a discrete quantum walk on a cyclic graph of $2^N$ nodes and a symmetric coin (see Ref.~\cite{Kempe_2003} for a comprehensive overview). The model includes two quantum variables, a qubit indicating the coin state and an array of qubits, $x$, representing the position of a walker along the circuit, i.e. an integer in $[0, 2^N-1]$; see the schematic description in Fig.~\ref{fig:increment_implementations}a.

The main quantum block of the algorithm is the increment function ($x\mathrel{+}=1 \pmod{2^N}$). We focus on a specific, yet flexible design for this quantum function, given as a series of multi-controlled X (MCX) operations\footnote{We note that there are various ways to model modular addition, e.g. by moving to Fourier space~\cite{Shakeel_2020,Razzoli_etal2024}.  However, we focus on the MCX based implementation as it highlights the rational of our approach. In fact, such an implementation was recently used to benchmark quantum hardware~\cite{Wadhia_2024}.}.  The MCX primitive has a variety of gate-level implementations, which differ in depth and number of auxiliaries~\cite{Barenco_etal1995,Maslov2016}. A high-level model (see Listings~\ref{lst:increment} and~\ref{lst:quantum_walk} in the Methods section) does not specify any such specific detail of the MCX implementation; this is determined dynamically by the synthesis engine given some global constraints for the full model. Panels (b) and (c) in Fig.~\ref{fig:increment_implementations} present two resulting quantum programs obtained by the synthesis engine for a circle with $2^5$ nodes, for two different optimization scenarios. We can see that when optimizing over the number of CX gates, the automated outcome includes two additional auxiliary qubits, allowing reduction of CX counts at the expense of the circuit's width. Both auxiliaries are used for the first MCX operation, then only one of them is re-used for the two consecutive operations, and both auxiliaries are finally reused for the controlled decrement operations.

\begin{figure}[h!]
\centering
\includegraphics[width=0.9\linewidth]{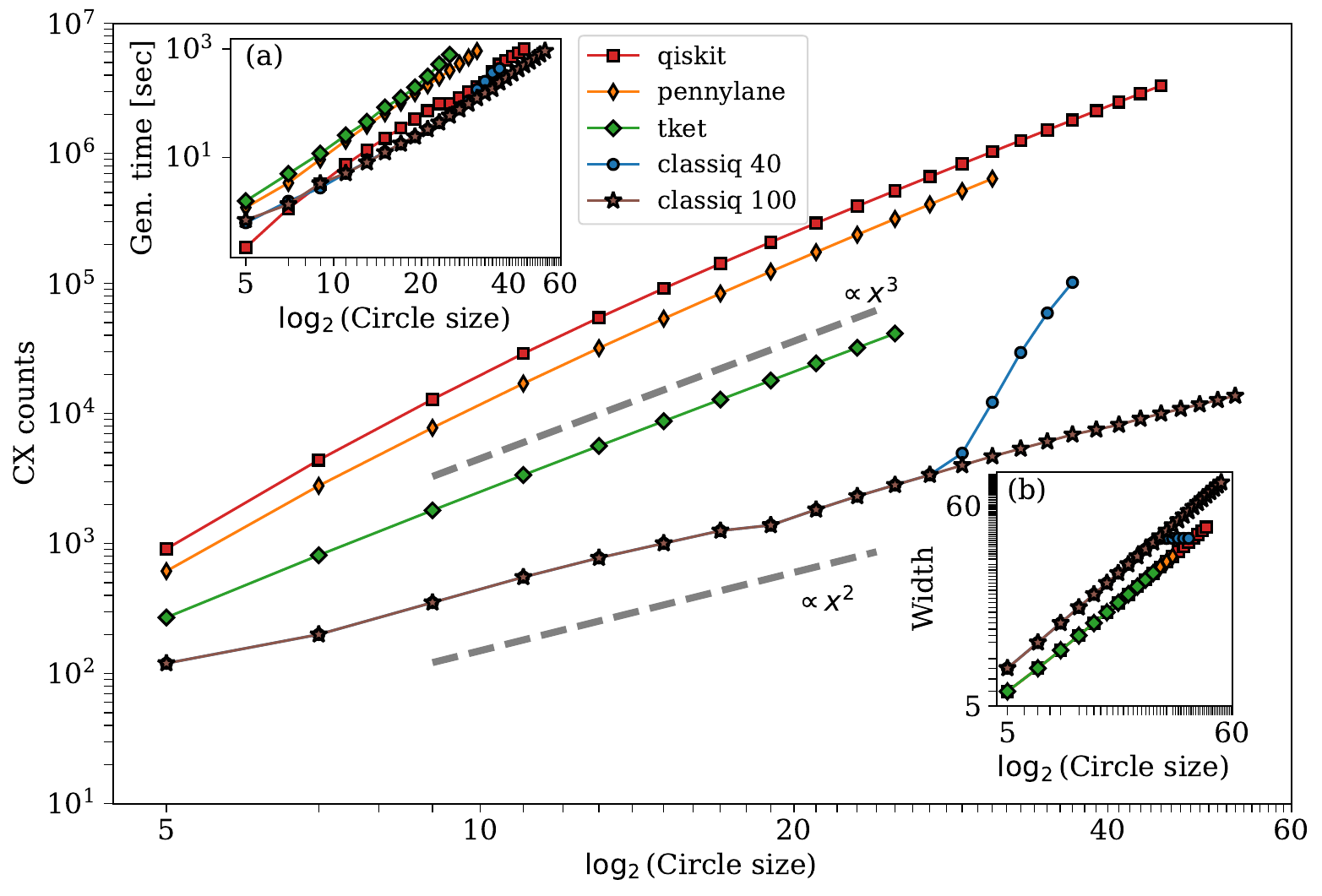}
\caption{
Large scale quantum walk programs on a circle of $2^N$ nodes. The Classiq curves are obtained by synthesizing with optimization over the CX counts and constraining the total width to 40 (blue dots) or 100 (brown stars). The main plot shows the total CX counts, and the two insets (a,b)  shows the total generation time and programs' width, respectively (we stop at compilation times of $\sim 10^3$ sec).
}
\label{fig:quantum_walk_scale}
\end{figure}

Trying to perform by hand the automation described in the last paragraph --- decisions on local implementations out of multiple possibilities and auxiliary reuse management --- is impractical. The effectiveness of this dynamic resource allocation becomes even more pronounced as we move to larger, more complex, and less structured models. In Fig.~\ref{fig:quantum_walk_scale} we run the same quantum walk model for different values of circle sizes $2^N$, synthesizing with optimization over the CX count and constraining with two different values of maximal width. Our in-house library contains flexible MCX implementations that, given a sufficient number of auxiliaries, can reach a CX count of $O(n_{\rm ctrl})$ with $n_{\rm ctrl}$ being the size of the control variable, resulting in a scaling of $O(N^2)$ of the quantum walk. From a software architecture perspective, any user can implement and provide any additional implementations of the MCX function (or any other function), and those will be taken into account by the synthesis engine just as the ones in our in-house library.

Fig.~\ref{fig:quantum_walk_scale} also shows a comparison to the best-of-the-art model implementations using several other quantum software tools. In all those cases, implementations have a fixed CX-count for any number of control qubits in the function, which scales quadratically with the number of control variable\footnote{In TKet~\cite{pytket_2020} this results in a cubic trend in the main plot of Fig.~\ref{fig:quantum_walk_scale}. The steeper curves of Qiskit~\cite{Qiskit2024} and Pennylane~\cite{Pennylane} for the CX-counts probably originates in inefficient compilation of the controlled-MCX function, not interpreting it as a larger MCX gate.}. We note that the libraries of these tools do have ad hoc, more efficient implementations for the MCX operation in terms of the CX gate count. However, when using these more efficient implementations, auxiliary qubits must be declared and treated as part of the function itself. Implementing the program with the compilers of these tools, which lack auxiliary awareness and management, results in a large circuit width because the auxiliaries of each function add up instead of being reused. In general terms, the underlying concept of all existing standard tools (including those that do include memory management~\cite{Silq2020, CudaQ, Qrisp2024}) is that they do not distinguish between the notion of a function and that of an implementation.  

We finally note that one can go a further step and consider multiple implementations based on different designs for the increment function, which do not originate in the flexibility of the MCX building block, e.g., by introducing QFT-based or ripple-carry adders. For the sake of simplicity, we have not incorporated this degree of freedom in the simple quantum-walk experiment, and we consider this effect when dealing with larger problems in the next section. 

\subsection{Quantum Singular Value Transform}
\label{sec:qsvt}

We now examine a more applicative and less-structured example, applying a Quantum Singular Value Transform (QSVT) on a sparse block-encoded matrix. QSVT is a quantum primitive that is at the core of many algorithms such as quantum linear solvers, Hamiltonian simulation, and amplitude loading (see Ref.~\cite{GrandUnification2021} and references therein). It supports a wide range of applications going from Finance~\cite{Stamatopoulos2024}, through Quantum Machine Learning~\cite{Gilyen_etal2019} to Plasma Physics~\cite{Toyoizumi_etal2024}. 

The implementation of QSVT is based on the block encoding of some matrix $A$ into a larger unitary matrix $U_A$, which can then be executed on quantum hardware. We block encode a simple $2^N\times 2^N$ matrix
\begin{equation}
A = \begin{pmatrix}
0   & 0   & 0   & \dots & 0   & 0 \\
0.5 & 0   & 0.5 & \dots & 0   & 0 \\
0   & 0.5 & 0   & 0.5   & \dots & 0 \\
\vdots & \vdots & \ddots   & \ddots     & \ddots & \vdots \\
0   & 0   & 0   & 0.5   & 0   & 0.5 \\
0   & 0   & 0   & 0     & 0.5 & 0
\end{pmatrix}.
\label{eq:Amatrix}
\end{equation}
Quantum speedup that is based on the QSVT methodology assumes that the problem, e.g., the linear equation to be solved or the Hamiltonian to be evolved, is sparse. We find that synthesizing a rather minimal use-case as Eq.~\eqref{eq:Amatrix} already shows significant improvement compared to state of the art compiler tools working without the synthesis principles.

\begin{figure}[h!]
\begin{minipage}{\hsize}
\centering
\includegraphics[width=0.8\linewidth]{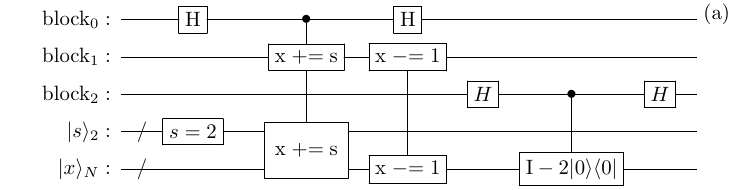}
\vspace{8pt}
\end{minipage}
\begin{minipage}{\hsize}
\includegraphics[width=0.5\linewidth]{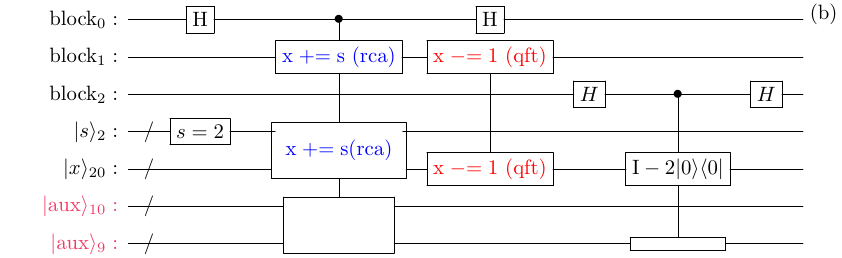}
\hfill
\includegraphics[width=0.5\linewidth]{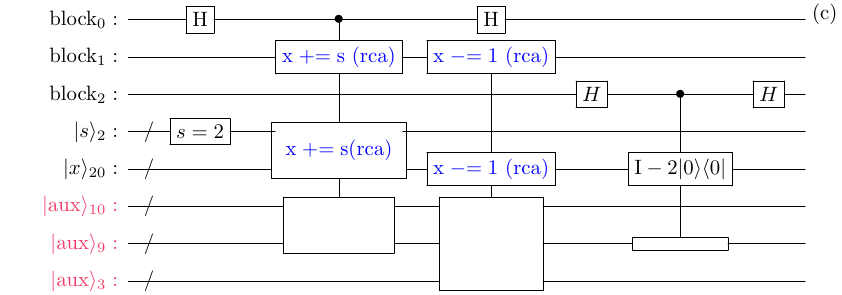}
\end{minipage}
\caption{Block encoding model for the matrix $A$ in Eq.~\eqref{eq:Amatrix}. (a): A schematic description of the model, the $I-2|0\rangle\langle 0|$ function is defined with  single-qubit gates and an MCX operation, see Methods. (b,c): Two different implementations for $N=20$ obtained by the synthesis engine when optimizing over CX-counts and constraining the width to 45 (b) or 90 (c) qubits. Different adders are chosen, QFT or ripple carry (rca), given the different constraints, resulting in CX-counts of 1480 and 842 for a 44 and 47 qubits circuit in panels (b) and (c), respectively.}
\label{fig:block_encoding}
\end{figure}

The block encoding of $A$ can be defined with quantum adders and reflection operations; see the schematic model in Fig.~\ref{fig:block_encoding}(a) and technical details in the Methods Section~\ref{sec:methods}. In particular, the encoding unitary is of size $2^{N+3}$, having 3 block qubits\footnote{In Fig.~\ref{fig:block_encoding} we add another quantum variable $|s\rangle$ for making the problem even less structured, supporting, e.g., superposition of block-encoded matrices.}. Our in-house library includes two different implementations of an in-place adder function on a quantum variable of size $M$: a QFT adder~\cite{Draper2000}, that uses no extra auxiliary qubits and has an $O(M^2)$ CX-counts, and a ripple-carry adder~\cite{Cuccaro200} that uses $M+1$ auxiliaries and has an $O(M)$ CX-counts. Fig.~\ref{fig:block_encoding} shows the implementations provided by the synthesis engine for the block encoding model with two different optimization scenarios.

\begin{figure}[h!]
\centering
\includegraphics[width=0.9\linewidth]{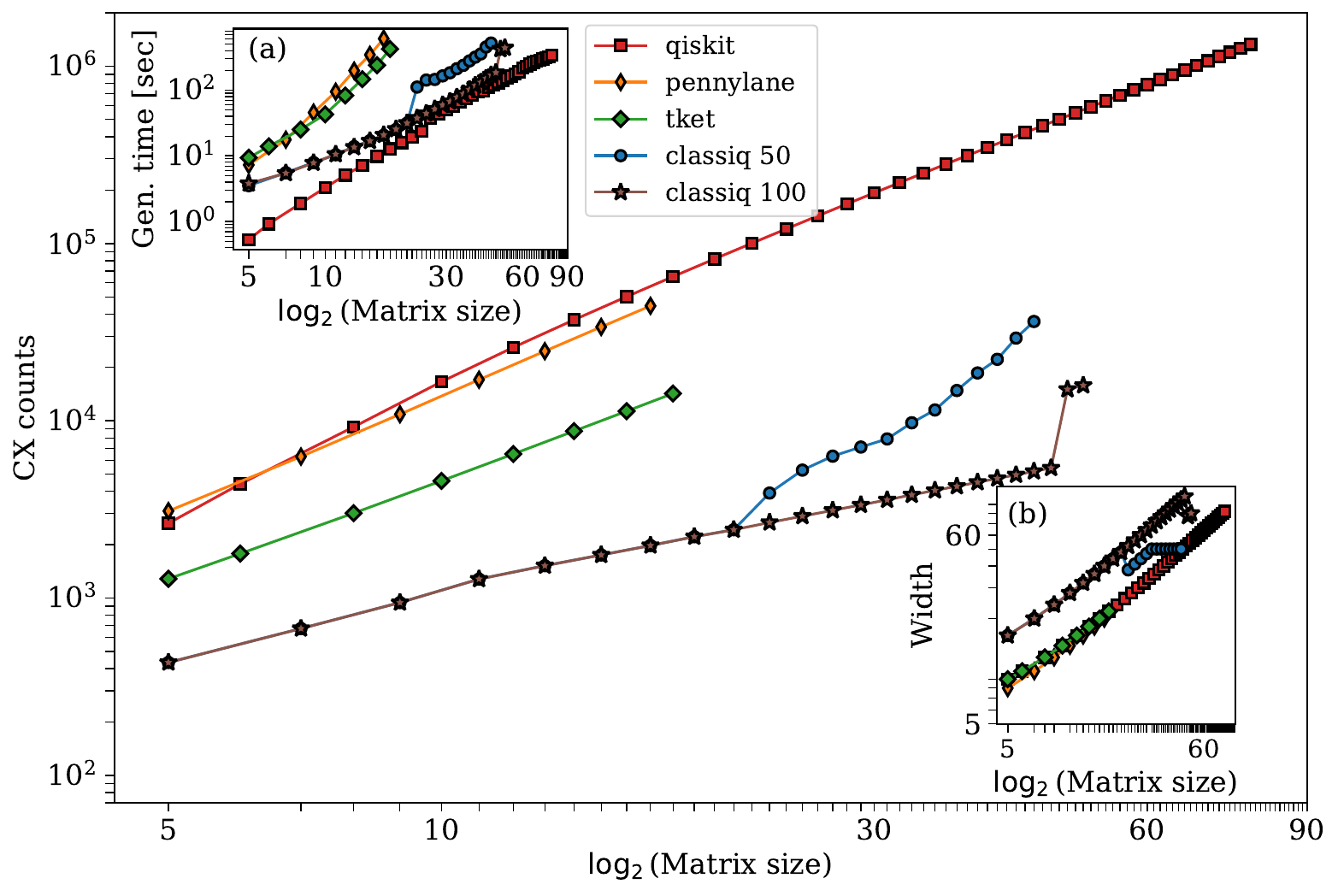}
\caption{QSVT for a polynomial of matrix $A$ in Eq.~\eqref{eq:Amatrix} for varying matrix sizes $2^N$. The main plot shows the total CX counts, and the two insets (a,b) shows  and total generation time and the quantum program width respectively (we stop at compilation times of $\sim 10^3$ sec). The Classiq curves are obtained by optimizing over the CX counts and constraining the total width to 50 (blue dots) or 100 (brown stars).
}
\label{fig:qsvt}
\end{figure}

Given a block encoding model, one can plug it into a QSVT routine to block encode a polynomial of degree $l$, $P_{l}(A)$~\cite{GrandUnification2021}. The  model corresponds to applications of $U_A$, its inverse, and rotations about the block states; see Methods. We generate a random odd polynomial of degree 3 and apply it to different values of the matrix size. Fig.~\ref{fig:qsvt}. shows our optimized results for CX counts, compared to state-of-the-art results from other software tools, originates from the flexibility of choosing efficient implementations of controlled operations and quantum adders.

\section{Conclusions}\label{sec:conclusions}
We have discussed the application of EDA synthesis methods to the field of quantum software and have shown clear results that support this approach as the robust, scalable and long-lasting way to approach quantum software development. One thing we have not covered in detail are the affect of different optimization functions and in particular fitness to a given hardware. For example, the graphs of Figs.~\ref{fig:quantum_walk_scale} and \ref{fig:qsvt} show results for which the synthesis engine optimizes for a minimal number of CX gates. They would look very different if the user decided to optimize for minimal width, providing much narrower circuits to the user. In addition, taking target hardware into account may have a crucial effect on the synthesis. For example, in fully connected hardware the synthesis engine may be more likely to implement an MCX gate with more CX gates than when the target is sparsely connected hardware, because in the latter case the cost of each CX gate is very large due to the need of swap gates for routing. We stress that such decisions on how to implement an MCX or other high-level functions cannot be taken by a transpiler which accepts a low-level implementation as input. Only true abstraction and true dynamic synthesis, as we employ in this paper, can make such crucial decisions. Lastly, we stress that the decision on how to implement a function may be different for each instance of the function in the program, depending on the resources available \textit{locally} for the function at this part of the program.

While we laid the foundations of the EDA-based synthesis concept and concretized it by an industry-grade software platform, there is still much to be done. A main item that is not yet fully implemented is the management of the accuracy resource. Specifically, automatic distribution of the global required algorithmic accuracy between the individual blocks in the program. Moreover, for noisy intermediate scale quantum (NISQ) target devices, the requirement is to decide on the accuracy of each block both from the overall required algorithmic accuracy and from the competition with the hardware noise. Further into the future are items related to the co-design and co-synthesis of the logical quantum program and the error-correction code it runs upon in fault-tolerant scenarios. We have applied EDA methods for determining the positioning of logical qubit patches in fault-tolerant schemes, such as to minimize two-qubit logical gate collisions and minimize overall program runtime. However, co-creating the logical program together with the error-correcting code is highly challenging as it involves the co-synthesis of coupled systems within two very different levels of abstraction.

Our technology adapts additional methods from EDA that were not described in this paper. These include the design of Qmod~\cite{Qmod_docs} as the high-level modeling language for quantum; debugging and analysis approaches to the quantum program coming out of the synthesis engine; modeling and synthesizing hybrid quantum-classical programs; visualization tools and graphical editors for quantum models and quantum programs; and nonfunctional modeling idioms for quantum. Examples of such idioms are requirements for levels of entanglement in specific parts of the program, phase difference between different parts of the program, and more. In all those areas, as well as in the main topic presented in this paper --- synthesis of quantum programs from high-level functional models --- there remains a huge amount of work to do in order to reach the level of a robust, scalable, reliable, usable, and fast software ready for addressing the demanding needs of quantum algorithm developers in a post-quantum-advantage industrial world.

\section{Methods}\label{sec:methods}
A detailed description of the synthesis engine is provided in the Supplementary Materials. Here we provide some technical details needed to perform the experiments of Sec.~\ref{sec:results}, while the corresponding code is available in a public repository~\cite{paper_code}.

\subsection{Numerical study details}

The results presented in the paper are obtained by modeling simple quantum models on various quantum platforms, using their Python SDK package for modeling and compilation per platform. The software versions used in this paper are: Classiq 0.60.0, Qiskit 1.2.4, PyTket 1.34.0, and PennyLane 0.39.0 with PennyLane-Catalyst compiler 0.9.0. For each platform, we use its quantum control operation, MCX function, basic single-qubit gate operations, and built-in quantum arithmetic operations (when available, otherwise we define it ourselves). In all cases, experts knowledgeable about the software tools made their best effort to apply the tools in the most reasonable way to achieve the task in question. The specific inputs to the tools are provided in Ref.~\cite{paper_code}.

In all platforms, we use the tool's compilation and transpilation abilities to get a final quantum program with only one-qubit gate and CX gate operations. This allows a fair comparison for the CX counts per example. All runtime experiments were performed on a single Apple M1 pro, all on the same computer with the same software installed and the same environmental conditions. During run-time, the software in question was the only significant application running on the computer.

\subsection{Quantum walk model}

We discuss in detail the quantum model presented in Sect.~\ref{sec:quantum_walk}. The model definition is accompanied with a Qmod code for highlighting the high-level description. A discrete quantum walk refers to the repetitions over "coin flipping" followed by a conditional step according to the coin state. For the circular graph, the coin operation is defined by a a Hadamard gate, represented by the $2\times 2$ matrix $H=\begin{pmatrix}
    1 & 1 \\
    1 & -1
\end{pmatrix}$.
This is a standard 1-qubit gate, available in all software tools and languages. A clockwise (counter clockwise) step is given by an increment (decrement) operation, referring to an in-place addition (subtraction) by 1, modulo the number of nodes $2^N$. We design the increment function as a cascade of MCX operations:
\begin{mdframed}[backgroundcolor=light-gray, roundcorner=10pt,leftmargin=1, rightmargin=1, innerleftmargin=15, innertopmargin=15,innerbottommargin=15, outerlinewidth=1, linecolor=light-gray]
\begin{lstlisting}[caption= An implemetation of an increment operation, label=lst:increment]
qfunc my_mcx(x: qnum, y: qbit) {
  control (x==((2**x.size)-1)) {
    X(y);
  }
}
qfunc increment(x: qbit[]) {
  repeat (i: x.len - 1) {
    my_mcx(x[0:(x.len - 1) - i],
           x[(x.len - 1) - i]);
  }
  X(x[0]);
}
\end{lstlisting}
\end{mdframed}
Then, a single walking step is defined via control and invert operations:
\begin{mdframed}[backgroundcolor=light-gray, roundcorner=10pt,leftmargin=1, rightmargin=1, innerleftmargin=15, innertopmargin=15,innerbottommargin=15, outerlinewidth=1, linecolor=light-gray]
\begin{lstlisting}[caption= An implemetation of a quantum walk on a circle, label=lst:quantum_walk]
qfunc single_step(coin: qbit, x: qnum)
{
  H(coin);
  control (coin == 0) {
    increment(x);
    invert {
      increment(x);
    }
  }
}
\end{lstlisting}
\end{mdframed}

The models in all other platforms are designed in the same fashion, using the platforms' building blocks such as quantum control and inversion operations~\cite{paper_code}.

\subsection{QSVT on the block-encoded $A$}

We block encode the matrix $A$ in Eq.~\eqref{eq:Amatrix} as a product of $A_0 = 0.5\sum^{2^N-2}_{i=0}(|i\rangle \langle i+1|+ 0.5|2^N-1\rangle \langle 0| + c.c$ and $A_1=I-|0\rangle\langle0|$, and apply block-encoding multiplication. The block encoding of $A_0$ is done with 2 block qubits, according to the procedure described in~\cite{Sunderhauf2024} for Toeplitz matrices [see Eq. (55) therein]. The matrix $A_1$ can be block encoded as a Linear Combination of Unitaries (LCU) for the unitaries $I$ and $I-2|0\rangle\langle0|$. The latter unitary is the reflection around zero operation appearing in Grover's Search algorithm. The Qmod code for block encoding $A$ is given in Listing~\ref{lst:be}.

\begin{mdframed}[backgroundcolor=light-gray, roundcorner=10pt,leftmargin=1, rightmargin=1, innerleftmargin=15, innertopmargin=15,innerbottommargin=15, outerlinewidth=1, linecolor=light-gray]
\begin{lstlisting}[caption= An implemetation of a block encoding A, label=lst:be]
qfunc be_amat0(data: qbit[], block: qbit[]) {
  del_qubit: qbit;
  select: qbit;
  packed: qnum<data.size + 1>;
  within {
    block -> {select, del_qubit};
    {data, del_qubit} -> packed;
    hadamard_transform(select);
  } apply {
    control (select == 0) {
      packed += 2;
    }
    packed += -1;
  }
}

qfunc be_projection(x: qbit[], aux: qbit) {
  within {
    H(aux);
  } apply {
    control (aux == 0) {
      reflect_about_zero(x);
    }
  }
}

qfunc be_amat(data: qbit[], block: qbit[]) {
  be_amat0(data, block[0:2]);
  be_projection(data, block[2]);
}
\end{lstlisting}
\end{mdframed}

The QSVT scheme for a polynomial with a given parity is a standard quantum primitive. It involves iterations over application of the block encoding unitary, its inverse, one-qubit rotations, and reflection operations (see Fig. 1 in Ref.~\cite{Gilyen_etal2019}). We use the classiq-library~\cite{classiq_library} functions to obtain the QSVT application, as well as for the reflection about zero unitary. The QSVT function in Classiq open library accepts a projection operator, given in Listing~\ref{lst:c_prog}.

\begin{mdframed}[backgroundcolor=light-gray, roundcorner=10pt,leftmargin=1, rightmargin=1, innerleftmargin=15, innertopmargin=15,innerbottommargin=15, outerlinewidth=1, linecolor=light-gray]
\begin{lstlisting}[caption= An implemetation of a reflection in the block space, label=lst:c_prog]
qfunc my_projector_controlled_phase(phase: real, block: qnum, aux: qbit) {
  control (block == 0) {
    X(aux);
  }
  RZ(phase, aux);
  control (block == 0) {
    X(aux);
  }
}
\end{lstlisting}
\end{mdframed}

\bmhead{Supplementary information}

The article is accompanied by a Supplemental Materials describing the synthesis engine. The code for obtaining using other quantum softwares and compilers is available in a public repository~\cite{paper_code}.

\section*{Declarations}

\subsection*{Conflict of interest/Competing interests}
The authors declare no competing interests.
\subsection*{Data and Code availability }
The datasets generated during and/or analyzed during the current study, and the code that generates them, are partially available in Ref.~\cite{paper_code}. The rest of the code is available from the corresponding author on reasonable request. 
\subsection*{Author contribution}
Classiq team developed, designed, and implemented the Synthesis engine, the Qmod language, and the classiq-library, and reviewed and adjusted the manuscript. T.G and Y.N, wrote the manuscript and I.R the Supplementary Materials. T.G ran the numerical experiments on the Synthesis engine and other quantum platforms.

\bibliography{classiq_bib}%

\newpage

\appendix

\section{Supplemental Material}

\subsection{Introduction and Purpose}
The Supplementary Materials provide in-depth technical details regarding the synthesis engine. At its core, the synthesis engine tackles a constrained optimization problem, converting a high-level model into a concrete implementation. This process requires specifying all the details not covered by the model while adhering to constraints and striving for the optimal value of the target function. The solution is achieved through a propagation-based constraint satisfaction problem (CSP) solver, enhanced by a branch-and-bound scheme for optimization.

This document serves a dual purpose. First, it aims to deliver comprehensive insights into the engine, allowing readers to grasp how the results presented in the main paper are derived. Second, it underscores the critical distinction between the synthesis approach and traditional linear, multi-pass compilation methods commonly utilized in leading quantum software tools. Importantly, many considerations discussed here emphasize the necessity of adopting a global perspective on the problem, rather than relying solely on local operations.

\subsection{Terminology}
\begin{itemize}
    \item Function Call Graph - A directed acyclic graph (DAG) that represents the dependencies among function calls in the input high-level model.
    \item Node - Within the context of the function call graph, we will use the term "node" interchangeably to refer to a function call.
    \item Composite Function - A function that includes inner functions. A composite function corresponds to a node that can be expanded into a graph.
    \item Flattening - The process of expanding a composite function.
    \item Implementation - 
    \begin{itemize}
        \item A black box, i.e., a low-level representation of a quantum circuit with no room for further decisions, that achieves the desired functionality.
        \item Note: The construction of these implementations may involve complex algorithms (such as trotterization, state preparation, etc.). The expression “no room for further decisions” indicates that once constructed, these implementations are fixed and do not allow for additional engine choices.
        
        \item For a composite function, an implementation consists of a specific set of inner function calls. This set is determined, leaving no opportunity to modify the functions to be called or the order in which they are called.
    \end{itemize}

    \item Topological Sorting - Linearization of a graph. A topological sort is a specific linearization.
    
    \item Domain - Denotes a discrete set or range of possible values for each function. For example, the domain representing the potential number of auxiliaries a function may use may be [0, 1, 2, 4].
\end{itemize}

\subsection{Classiq’s Synthesis Engine}

The engine’s objective is resource allocation. Allocation is done by by solving a constraint satisfaction problem over the resource variables (width, depth, different gate counts, and more) of the overall quantum program to be created from the model. An optional optimization goal over those variables may be added as well.

The outcome quantum program is at the level of gate-level quantum circuits that can be expressed in any common format such as QASM, Q{\#}, cirq, and more.

In more detail, the engine receives the following inputs:

\begin{itemize}
    \item A function call graph representing the input model
    \begin{itemize}
        \item The call graph may be more abstract then the single-function level due to multiple possible implementations composite functions.
    \end{itemize}
    \item Information on whether each function requires qubits from the pool or releases qubits when terminated.
    \item Domains representing possible values associated with the different implementations of the functions: widths, depths, and gate counts. Values for different implementations may be calculated by the engine on the fly through utility functions, or be accepted from the user implementing the functions.
\end{itemize}

The outputs of the engine are
\begin{itemize}
    \item A specific choice of resource values for each function.
    \item Relations between the functions: a topological sort for the graph together with information on qubit reuse.
\end{itemize}

The engine generates its output by solving a CSP problem. Depending on the problem size and complexity, and on the available runtime, this can be done either by running a full CSP flow or by separating the problem into smaller, weakly coupled CSP’s, as discussed below. 

Note that the engine does not end with a concrete implementation for each function, but only with a concrete set of resources determined for each function. The actual implementations will be determined downstream from the engine, according to further heuristics, and between all implementations fitting the determined resources.

\subsection{Note about synthesis}

Our current definition differs from the common definition of unitary synthesis found in academic papers or the open-source community, which usually refers to generating a gate-level quantum circuit out of a given unitary. Here the term is generalized to finding a concrete gate-level quantum program which implements an abstract functional high-level model. Thus ‘synthesis’ in our context refers to the entire scheme doing this task, and not to a specific decomposition algorithm. This usage is in line with the common usage of the word synthesis in the electronic design automation (EDA) domain.

\subsection{The CSP Solver}
The engine's basic CSP flow is a simple implementation of the backtracking algorithm~\cite{FREUDER200613}. It creates a tree of possible values representing function values, the topological sort, and qubit reuse options. Then, it traverses these decisions through a depth-first-search (DFS) over the decision tree.

Each decision corresponds to assigning a value from the domain, which shrinks the domain to a singleton. If a specific set of assignments violates the constraints, the engine backtracks. 

We also implement pruning mechanisms that simplify the decision tree at any point in the search. this will be explained in detail below.

\subsubsection{Decisions}

We here specify the possible decisions that can be made by the engine. The assignment is incremental - the engine never assign multiple values at once. 

The order of the assignments is strict and important because the engine can’t decide on qubit reuse before  the topological sort and the number of auxiliary qubits corresponding to the previous functions are decided.

\paragraph{Decision Points in the Simple Case - No Multiple Implementation Composites}

\begin{itemize}
    \item Node fail - a placeholder to wrap up the data structures and backtrack to the previous node choice.
    \item Next node - move to next node in the topological sort.
    \item Node values - number of auxiliary qubits, depth, and gate count, according to constraints.
    \begin{itemize}
        \item Heuristics: If the engine optimizes over a constraint, we decide on the value associated with the constraint. For example, under CX gate count optimization, the first decision will correspond to the number of CX gates.
    \end{itemize}
    \item Reuse qubit count and reuse options.
    \item Node Done - a placeholder to wrap up the data structures and go to the next decision sequence.
\end{itemize}
The decision tree is implemented as a stack:
\begin{lstlisting}
push "node fail"
push [next_node_1, next_node_2, next_node_3]
pop next_node_1
push [num_aux_1, num_aux_2]
pop num_aux_1
push [depth_1]
pop [depth_1]
push [reuse_1, reuse_2] # The engine splits this part into two, but we combine them here for simplicity.
pop reuse_1
push "node done"
pop "node done"

push "node fail"
push [next_node_4]
pop next_node_4
# pruning - cannot continue
pop "node fail"

pop reuse_2 # from the previous step
# pruning - cannot continue
pop num_aux_2
push [depth_2]
# pruning - cannot continue
pop next_node_2
...
\end{lstlisting}

\paragraph{Decision Points in the Complex Case - Multiple Implementation Composites}

Multiple implementation composite functions introduce an additional complexity, where the function call graph is altered as part of the CSP process. A composite node is expanded to one of its implementations, which turns the node into a subgraph.

\begin{itemize}
    \item Node Fail - same as in the simple case
    \item Next node-  move to next node in the topological sort - it could now be a composite node. 
    \begin{itemize}
        \item Note that after the composite node has been expanded, it is removed from the topological sort.
    \end{itemize}
    \item Node values - the same as in the simple case
    \begin{itemize}
        \item Theoretically, we can estimate the values for composite functions to rule out certain implementations in advance. Practically, we don’t do it due to design complexity.
    \end{itemize}
\end{itemize}
These two following decisions apply to composite functions only:
\begin{itemize}
    \item Logic flow - the composites implementations. 
    \begin{itemize}
        \item After the expansion of a specific implementation, the function call graph contains the composite inner nodes and the other nodes. 
        \item We now go back to the beginning of the decision sequence - “Node Fail”, next node, etc. 
        \item Note that even though we expanded a composite function, the next node in the topological sort could be one that is not part of the composite function
        \item Obviously, this flow can be done recursively in case of a multiple implementation composite inside a multiple implementation composite.
    \end{itemize}
    \item Collapse - this corresponds to ‘Node Fail’ for the specific node values of the composite function
\end{itemize}
The following two decisions apply to non-composite (“elementary”) functions only:
\begin{itemize}
    \item Reuse qubit count and reuse options.
    \item Node Done - same as in the simple case.
\end{itemize}

Note that one issue with the above method is that it could miss potential topological sorts. If a composite function bar depends on another function foo, it doesn’t dictate that all of bar’s inner components depend on foo. However, the sequential nature of our current engine forces every inner node of bar to be positioned after foo in the topological sorting. Testing has shown that this issue does not create severe impact on the results for now.

The decision tree now looks as follows:
\begin{lstlisting}
    push "node fail"
push [next_node_1, next_node_2] # next_node_1 is a composite
pop next_node_1
push [num_aux_1]
pop num_aux_1
push "collapse"
push [logic_flow_1, logic_flow_2]
pop logic_flow_1

push "node fail"
push [next_node_2, next_node_3, next_node_4]
... # all pruned
pop "node fail"

pop logic_flow_2
push "node fail"
push [next_node_5]
pop next_node_5
push [num_aux_1, num_aux_2]
... # all pruned
pop "node fail"

pop "collapse"
pop next_node_2
...
\end{lstlisting}

\subsubsection{Optimization - Branch and Bound}
We apply a simple branch and bound optimization scheme: We start with the first decision sequence and calculate the corresponding value (e.g., depth = 500 when minimizing depth). 
 Then, we set a constraint specifying a need for a smaller value (i.e., depth $<500$), and continue to solve the resulting CSP problem. The same happens each time a full configuration which satisfies the constraints is found. The procedure ends when the engine does not find a satisfying configuration. Then the last satisfying configuration is returned as the best solution found

\subsection{Heuristic-Based Guesses - Solving Strategies}
Warm starting: Before initiating the full CSP solver, we start with initial guesses. Those are implemented as a degenerate case of the CSP flow. We traverse the decision tree and perform an assignment from a single choice that is pre-calculated according to the solving strategy.

Apart from serving as a guess, these strategies can lower the bound for branch and bound optimization, allowing better pruning before running the full CSP flow and thus potentially reaching the optimized solution faster.

The different strategies are given as follows:

\begin{table}[]
    \centering
    \normalsize
    \begin{tabular}{|p{4cm}|p{4cm}|p{4cm}|}
        {\bf Strategy}   & {\bf Is Applied When} & {\bf Notes}\\
        \hline
        Greedy reuse:
        \begin{itemize}
            \item Flatten all composites
            \item Topological sort heuristically attempts to maximize reuse possibilities by arithmetic uncomputation.
            \item Reuse the maximum number of qubits possible.
            \item Number of auxiliaries - closest to the domain average. 
        \end{itemize}
& The model is unconstrained and no optimization function is provided &  
            In such cases it may be thought that any solution is acceptable. However, there is an implied requirement here - even if the user did not specify it explicitly - that the program must make sense in the eyes of an expert, e.g., be dense enough, of not use unnecessary resources. In other words, the program should aim to be not outside the Pareto curve\\
            \hline
            Minimal width
            \begin{itemize}
                \item Choose the next node according to heuristics set to reduce width.
                \item Reuse the maximum number of qubits possible.
                \item For the number of auxiliaries value - choose the minimum
            \end{itemize}& 
            The model includes a width constraint or optimization.&
            Use the greedy reuse to dictate the topological sort.\\
            \hline
            Minimal depth
            \begin{itemize}
                \item Choose the next node randomly
                \item Choose the implementation with the minimal depth
                \item Don’t reuse qubits at all
            \end{itemize}&
            The model includes a depth constraint or optimization.
            & Use the greedy reuse to dictate the topological sort. Note that topological sort doesn’t affect depth. Only when we reuse qubits we could affect the depth.
            
            Reuse qubits if there exists a dependency between the functions (which means that the depth is not affected)\\
            \hline
    \end{tabular}
    \label{tab:table1}
\end{table}

\begin{table}[]
    \centering
    \normalsize
    \begin{tabular}{|p{4cm}|p{4cm}|p{4cm}|}
    \hline
            Minimal depth with minimal reuse. Same as minimal depth strategy, but reuses the smallest amount of qubits possible to satisfy width constraints &
            The model includes both depth and width constraints and optimization.
            & As for minimal depth strategy \\
            \hline
            Random
            
            Everything is chosen randomly &
            The model is constrained by arbitrary constraints
            &\\
            \hline
    \end{tabular}
    \caption{Strategies of the synthesis engine.}
    \label{tab:table_2}
\end{table}

\subsection{Graph Reducer}
The “graph reducer” is applied before every strategy, except when greedy reuse is used, because the latter dictates a single topological sort in advance. 

\begin{figure}[h!]
\centering
\includegraphics[width=0.9\linewidth]{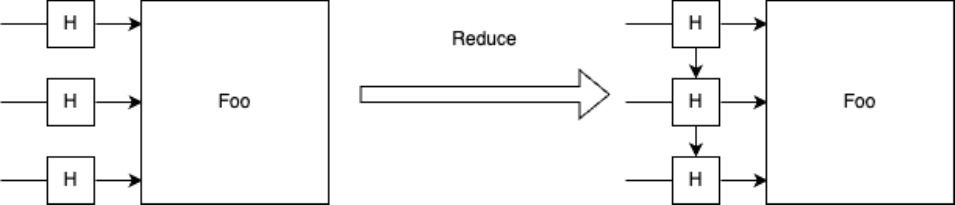}
\caption{Graph reducer example.}
\label{fig:diagram}
\end{figure}

The role of the graph reducer is to add artificial edges between graph nodes to create additional relations. Why is it important? Consider a circuit where the quantum state is initialized with ten Hadamard gates. Obviously, all 10! permutations of ordering the gates are equivalent, and the CSP solver needs to test only one (see Fig.~\ref{fig:diagram}).

Note that this flow is only relevant for node selection. Since we aim to use the greedy reuse pass for every solving strategy, we should use the reducer only during the complete CSP solution stage.

\subsection{Domain Initialization}
Domains are initialized before each solving strategy. They can be truncated according to the overall constraints to significantly reduce the CSP search space. They are also reduced to singletons before the heuristic-based guesses.
For example: 

\begin{itemize}
    \item If the only optimization criteria is on the CX gate count, the domain of each function is reduced to a singleton that minimizes the CX gate count.
    \item If the solving strategy is minimal depth, all depth domains are reduced to a singleton.
\end{itemize}

Domain reduction also applies when multiple constraints are present if no trade-offs exist between values (also referred to as domination). 

Note that such strategies are not mathematically complete, as an optimized solution may lie beyond those decisions. However, in practice we have not encountered such pathological cases.

\subsection{Pruning - Constraint Propagation}

The constraint propagation is the heart of the CSP engine and what allows it to be effective. It can tell us in advance which branches of the decision tree are not worth visiting and prune them~\cite{FREUDER200613}, saving us exponential-time traversal.

\subsubsection{The Arc Consistency Pass}
Arc consistency is used to truncate the domains of the number of auxiliaries, depth, and gate count. 

It does not truncate the reuse options, which the engine enforces locally. We will discuss this in the next section.

\paragraph{Single Variable and Constraint}

It is best to start with a simple example to present arc consistency. 
\begin{itemize}
    \item Suppose that there exist two variables, $x_1$ and $x_2$, each of which can take three possible values: $\{1500, 2000, 3000\}$.
    \item Now, we add a constraint: $x1 + x2 \leq 3500$. 
    \item We can immediately tell that 3000 is an illegal value for $x_1$ because the minimal value for $x_2$ is 1500, and vice versa. Therefore, the new possible values for these variables are $\{1500, 2000\}$. The CSP domain is shrunk. 
    \item However, we can’t rule out the value 2000 individually because the assignments $x_1=1500$, $x_2=2000$, or $x_1=2000$, $x_2=1500$, are still legal. Arc consistency is an optimistic method, meaning we only rule out worst-case scenarios.
    \item Suppose we assign $x_1=2000$ now. Then, after running the arc consistency again, we deduce that $x_2$ can only take the value 1500.
    \item Note that if the constraint was $x1 + x2 \leq 2500$, we would have reached an empty domain. This is referred to as inconsistency.
\end{itemize}
Simple as it is, this is how we perform pruning for the engine’s constraints:
\begin{itemize}
    \item The gate count is estimated conservatively. The gate count of each function is summed to estimate the total gate count and perform constraint propagation. Conservative here means that we don’t perform pattern lookups for possible gate cancellations at this stage.
    \item The depth is estimated conservatively like the gate count. It is calculated over the function call graph, and the path resulting in the maximal depth is considered.
    \item The number of auxiliaries is estimated exactly from the width constraint. 
    \begin{itemize}
        \item If a topological sort is given, the arc consistency pass is straightforward. It might be unintuitive, but a decision on the number of auxiliaries does not affect the number of auxiliary domains of the rest. Only the topological sort does. This behavior is explained by the fact that qubits that haven’t been reused before could be reused later. Recall that arc consistency is optimistic.
        \item It is more complex if we don’t know the topological sort in advance. In this case we perform a look-ahead into all possible qubit-releasing nodes and subtract the qubits required to reach said nodes. We aim to be optimistic in order to not prune legal solutions, but under this constraint we aim to be  as pessimistic as possible.
    \end{itemize}
\end{itemize}

\paragraph{Multiple Variables}
The introduction of new variables imposes additional constraints that enforce the relation between them. For example, a function can take the values $\{0, 1, 2\}$ for the number of auxiliaries and $\{3500, 500, 100\}$ for the CX gate count. The additional constraint is the correspondence between the variables due to their belonging to a single implementation: 0 auxiliaries correspond to 1500 cx gates, 1 auxiliary to 500, and 2 to 250.

We can now envision the arc consistency as running in two perpendicular axes: the variable axis, propagating the global constraint to the specific variables of all nodes (e.g., depth $<8000$ ), and the function axis, propagating the relation between the variables.

The arc consistency pass in this case runs iteratively and could be performed multiple times until convergence.

\paragraph{Skipping Arc Consistency for Performance}
The engine can predict whether applying an arc consistency pass results in the domains remaining intact, therefore saving the need to run the pass.

First, we note that arc consistency should never be skipped after finding a solution during optimization. The branch and bound method reduces the constraint value, and we have to check that the remaining branches satisfy the new constraint.

Otherwise, the engine skips arc consistency in the following cases:
\begin{itemize}
    \item After the reuse qubit count decision - since the engine hasn’t applied the actual reuse yet. The latter is done in the next decision, reuse options.
    \item The decision assigns a value while the domain is already a singleton - i.e., the domains have not been changed.
    \item The decision is reuse option, and either one or more of the following:
    \begin{itemize}
        \item The depth is not limited. This is because reuse only affects the depth of the future node choices and not their number of auxiliaries. Again, this might be unintuitive.
        \item The depth was not modified after reuse. 
        \begin{itemize}
            \item This probably means that there already exists a dependency between the releasing function and the reusing function.
        \end{itemize}
        \item The possible decisions form an empty set. This is a trivial special case of the previous condition.
    \end{itemize}
\end{itemize}
For multiple implementation composites, we skip arc consistency in the following case:
\begin{itemize}
    \item After Logic flow - because we don’t provide new information to the engine when we pick an implementation for a composite function.
\end{itemize}

\paragraph{Caching Domains for Backtracking}
To backtrack due to inconsistency, we need to restore the previous domain values before they were truncated due to an engine’s decision or an arc consistency pass. 

However, keeping the domain values for each node, variable, and engine step is expensive and requires many deep copy operations. We apply a caching mechanism that tracks the domains to be truncated, and they are the only ones whose previous values are stored for future backtracking.

\subsection{Handling Reuse Options}
The engine currently does not distinguish between different zero-valued qubits apart from their possible depth accumulation.

We store the auxiliary qubit depth. If a node reuses an auxiliary, the overall depth associated with this node is the maximum between: 
\begin{itemize}
    \item The auxiliary depth combined with the node depth.
    \item The accumulated node depth, regardless of the auxiliary.

\end{itemize}

\paragraph{Depth Domination Checks (Pareto Consistency)}
Consider the following example:

Suppose the engine decides to use two qubits. Now, it needs to choose an option from the following auxiliary depths: $\{800, 440, 150\}$. The options are $\{800, 440\}$, $\{440, 150\}$, $\{800, 150\}$.

The latter, however, can be ruled out. Why? Because the choice of 800 has already done the damage of increasing the depth and it is always better to keep 150 for late use than 440. The choice $\{800, 440\}$ dominates $\{800, 150\}$. This domination check allows us to reduce the multi-qubit reuse search space from exponential complexity in the number of reuse qubits to linear, up to a small overhead of sorting the array.

\paragraph{Enforcing Qubit Reuse Locally}
We can calculate the minimal number of qubits that need to be reused when reaching a node by summing over the qubits consumed by the predecessors and subtracting them from the overall width limit. We then compare it against the maximal number, which is the maximum between the available qubits to reuse and the node's qubit requirement. 

Because of the arc consistency, the full CSP flow will always result in the minimal number of qubits for reuse being smaller than the maximal number. It is guaranteed from the auxiliary qubit calculation. This is slightly unintuitive and highlights the complex dependency between auxiliary qubits and reuse options.

For strategies of initial guesses, however, the accumulated topological sort could be unfit. This will result in the minimal number of qubits required being greater than the number of available qubits for reuse.

\subsection{Context Objects}

Context objects store information relevant to the engine components and holds their current state during the CSP tree traversal.
The context objects are listed as follows:
\begin{itemize}
    \item Model context - provides information on individual nodes and the specific connections between between them.
    \item Graph context - stores the graph (constructed directly from the model), optimized data structures for graph algorithms, and the accumulated topological sort. It contains methods to compute the next possible node choices.
    \item Domain context - stores the domain values of each node at each engine’s step (with methods to avoid unnecessary copies).
    \item Propagator context - contains information relevant to the propagators, like pre-calculated and optimized data structures. The engine distills the required data from the propagator context and passes it to the constraint propagator during the arc consistency pass. A propagator context is formed for each constraint.
    \item Auxiliaries context - contains information on the qubits available for reuse. More specifically, it stores the depth of each reuse option.
\end{itemize}
The order of updating the context objects is important, and it is given by the order of appearance in the above list.

\subsection{Design Considerations and Compatibility with Other Approaches}

The details of the ideal solution architecture of the above scheme requires broad considerations. Specifically, we believe that an holistic approach is required to deal with program creation, compilation, resource estimation and optimization, and execution. This flow can be envisioned basically as a “model” and a series of “operations”, or “passes,” that transform a legal model into another legal model. 

A CSP solver could be treated as a special case of operations and passes. So are the compilation passes that predate the CSP solver, as well as future operations like variable assignments before execution and searching for further possible optimizations. 

A holistic design allows us to augment the CSP solving with other mechanisms. A disadvantage of CSP solving is that it requires users to specify values (e.g., depth $<10000$) while they are generally interested in targeting a broader behavior (for example, reasonable depth and gate count while staying within the qubit count limit of the hardware). 

We can envision the flow as three main different representations, which could be mixed. Each representation could be split further into more intermediate representations. The representations could be embedded into a framework like MLIR. Currently, only the variable semantics representation is implemented as a front end to the synthesis engine. The other two representations are listed for completeness and as indication of future work.

\paragraph{Variable Semantics (User-facing Model)}
This representation is the highest level one and reflects how the users write the model. After compilation procedures, it is transformed into an abstract syntax tree that is passed to the synthesis.

Note the select operation that acts as a directive for the synthesis to perform implementation choices. This representation is the easiest for the users to write programs in. We don’t expect the users to interact with other representations. It is also the simplest for composite function flattening, which is simply function inlining.  It can also host some basic optimization passes which include dealing with special cases or control, power, inversion, in-place arithmetic, and other operations. Below is an example in Qmod~\cite{Qmod_docs} demonstrating usage with variable semantics:

\begin{lstlisting}
@qfunc
def foo_1_implementation_1(q: QArray) -> None:
  ...
  
@qfunc
def foo_1_implementation_2(q: QArray) -> None:
  ...

@qfunc
def foo_1(q: QArray) -> None:
  select(
    [foo_1_implementation_1(q), foo_1_implementation_2(q)]
  )

@qfunc
def foo_2(q: QArray) -> None:
  ...

@qfunc
def main() -> None:
  allocate(num_qubits, q)
  foo_1(q)
  foo_2(q)
\end{lstlisting}

\paragraph{Wire Semantics - The Single Statement Assignment (SSA) Graph Representation}
Here, we replace the variables with wires, forming a function call graph.

Note that the conversion between wires and variables is pretty straightforward (though there’s an added complication in the cases of slicing, struct field access, etc.~\cite{Qmod_docs}). This representation is better suited for running graph-based routines like topological sort and depth calculations, as well as more complex optimizations like pattern lookups.

\begin{lstlisting}
@qfunc
def foo_1_implementation_1(w: Wire) -> Wire:
  ...
  
@qfunc
def foo_1_implementation_2(w: Wire) -> Wire:
  ...

@qfunc
def foo_1(w: Wire) -> Wire:
  return select(
    [foo_1_implementation_1(w), foo_1_implementation_2(w)]
  )

@qfunc
def foo_2(w: Wire) -> Wire:
  ...

@qfunc
def main() -> None:
  allocate(num_qubits, q)
  w_1 = var_to_wire(q)
  
  w_2 = foo_1(w_1)
  w_3 = foo_2(w_2)

  q = wire_to_var(w_3) # collect to perform execution measurements, etc.
\end{lstlisting}

\paragraph{The Solver representation}
This representation contains information about the resources consumed by each function. It is useful for solving resource constraints or optimization problems (via CSP or other methods). 

Note that this representation extends the other two. The select operation still contains the implementations to construct the circuit back after finding a solution. It makes sense to use the SSA graph representation as a basis because most resource estimation procedures require information about the dependency graph. However, it is not mandatory. For example, when expanding a composite function, it could be more convenient to transform all the way back to variable semantics, where a composite function expansion is a simple inlining. 

We also must note that if we choose an implementation with auxiliaries, the model extend foo\_1’s signature to include this auxiliary variable. This highlights that flexible transformation routines allow much better reasoning about the synthesis process rather than an engine with a fully rigid pipeline.

\begin{lstlisting}
@qfunc
def foo_1_implementation_1(w: Wire) -> Wire:
  ...  # a black box
  
@qfunc
def foo_1_implementation_2(w: Wire) -> Wire:
  ...  # a black box

@qfunc
def foo_1(w: Wire) -> Wire:
  return select(
    {
      "num_auxiliaries": [0, 4],
      "depth": [1500, 300],
      "cx": [800, 115]
    },
    [
      foo_1_implementation_1(w), 
      foo_1_implementation_2(w)
    ]
  )

@qfunc
def foo_2(w: Wire) -> Wire:
  ...

@qfunc
def main() -> None:
  allocate(num_qubits, q)
  w_1 = var_to_wire(q)
  
  w_2 = foo_1(w_1)
  w_3 = foo_2(w_2)

  q = wire_to_var(w_3) # collect to perform execution measurements, etc.
\end{lstlisting}
The model is expandable to implement functions with a resource estimation function based on a given parameter on an infinite domain:
\begin{lstlisting}
@qfunc
def foo_1(w: Wire) -> Wire:
  alpha = param(range=(0, 1))
  return select(
    alpha,
    {
      "error_bound": alpha,
      "depth": depth(alpha),  # some function, or a runtime call to a resource estimator to evaluate
      "cx": cx(alpha),   # some function, or a runtime call to a resource estimator to evaluate
    },
    foo_1_implementations(alpha, w)
  )    
\end{lstlisting}

\paragraph{Constructing The Solution}
Both the Qmod and the SSA representation can be used to construct the solution. A valid solution requires that the select operations have been reduced to a single statement and that there are no allocate or free operations inside inner functions. The latter condition means that the reuse options were fully assigned.

\end{document}